\documentclass[english,amsmath,amssymb,superscriptaddress,prl,nofootinbib]{revtex4-1}
\usepackage[T1]{fontenc}
\usepackage[latin9]{inputenc}
\usepackage{geometry}
\usepackage{color}
\definecolor{note_fontcolor}{rgb}{0.80078125, 0.80078125, 0.80078125}
\usepackage{verbatim}
\usepackage{float}
\usepackage{units}
\usepackage{textcomp}
\usepackage{amsmath}
\usepackage{amssymb}
\usepackage{esint}
\PassOptionsToPackage{normalem}{ulem}
\usepackage{ulem}

\makeatletter

\providecommand{\tabularnewline}{\\}
\newenvironment{lyxgreyedout}
  {\textcolor{note_fontcolor}\bgroup\ignorespaces}
  {\ignorespacesafterend\egroup}

\@ifundefined{textcolor}{}
{%
 \definecolor{BLACK}{gray}{0}
 \definecolor{WHITE}{gray}{1}
 \definecolor{RED}{rgb}{1,0,0}
 \definecolor{GREEN}{rgb}{0,1,0}
 \definecolor{BLUE}{rgb}{0,0,1}
 \definecolor{CYAN}{cmyk}{1,0,0,0}
 \definecolor{MAGENTA}{cmyk}{0,1,0,0}
 \definecolor{YELLOW}{cmyk}{0,0,1,0}
 }

\usepackage{amsthm}
\usepackage{calrsfs}
\usepackage[T3,T1]{fontenc}
\usepackage{mathdesign}
\usepackage{textcomp}
\usepackage{fge}
\makeatletter 
\newcommand*{\shifttext}[2]{%
  \settowidth{\@tempdima}{#2}%
  \makebox[\@tempdima]{\hspace*{#1}#2}%
}
\makeatother
\DeclareMathAlphabet{\mathpzc}{OT1}{pzc}{m}{it}
\DeclareSymbolFont{tipa}{T3}{cmr}{m}{n}
\DeclareMathAccent{\invbreve}{\mathalpha}{tipa}{16}
\newcommand{\alphaRealFirstMin}{\overset{\raisebox{0pt}{$\tiny{\mathrel{\ooalign{\hss$\smile$\hss\cr\shifttext{3.5pt}{\raisebox{-0.75pt}{1}}}}}$}}{\alpha_\Re}}

\makeatother

\usepackage{babel}
\begin{document}

\title{Optimal representation of the bath response function \& fast calculation
of influence functional coefficients in open quantum systems with
BATHFIT 1}

\author{Nikesh S. Dattani}

\email{nike.dattani@chem.ox.ac.uk}

\selectlanguage{english}%

\affiliation{Physical and Theoretical Chemistry Laboratory, Department of Chemistry,
University of Oxford, Oxford, OX1 3QZ, UK}

\author{David M. Wilkins}

\email{david.wilkins@seh.ox.ac.uk}

\selectlanguage{english}%

\affiliation{Physical and Theoretical Chemistry Laboratory, Department of Chemistry,
University of Oxford, Oxford, OX1 3QZ, UK}

\author{Felix A. Pollock}

\email{felix.pollock@physics.ox.ac.uk}

\selectlanguage{english}%

\affiliation{Clarendon Laboratory, Department of Physics, University of Oxford,
Oxford, OX1 3PU, UK}

\date{$\today$}
\begin{abstract}
Today's most popular techniques for accurately calculating the dynamics
of the reduced density operator in an open quantum system, either
require, or gain great computational benefits, from representing the
bath response function $\alpha(t)$ in the form $\alpha(t)=\sum_{\bar{K}}^{K}p_{\bar{K}}e^{\Omega_{\bar{K}}t}$.
For some of these techniques, the number of terms $K$ in the series
plays the lead role in the computational cost of the calculation,
and is therefore often a limiting factor in simulating open quantum
system dynamics. We present an open source MATLAB program called BATHFIT
1, whose input is any spectral distribution function $J(\omega)$
or bath response function, and whose output attempts to be the set
of parameters $\{p_{\bar{K}},\Omega_{\bar{K}}\}_{\bar{K}=1}^{K}$
such that for a given value of $K$, the series $\sum_{\bar{K}}^{K}p_{\bar{K}}e^{\Omega_{\bar{K}}t}$
is as close as possible to $\alpha(t)$. This should allow the user
to represent $\alpha(t)$ as accurately as possible with as few parameters
as possible. The program executes non-linear least squares fitting,
and for a very wide variety of forms for the spectral distribution
function, competent starting values are used for these fits. For most
forms of $J(\omega)$, these starting values, and the exact $\alpha(t)$
corresponding to the given $J(\omega)$, are calculated using the
recent Padé decomposition technique - therefore this program can also
be used to merely implement the Padé decomposition for these spectral
distribution functions; and it can also be used just to efficiently
and accurately calculate $\alpha(t)$ for any given $J(\omega)$.
The program also gives the $J(\omega)$ corresponding to a given $\alpha(t)$,
which may allow one to assess the quality (in the $\omega{\rm -domain})$
of a representation of $\alpha(t)$ being used. Finally, the program
can calculate the discretized influence functional coefficients for
any $J(\omega)$, and this is computed very efficiently for most forms
of $J(\omega)$ by implementing the recent technique published in
\cite{Dattani2012b}. We also provide a Mathematica program that can
perform this last calculation, along with calculating an analytic
form for these discretized influence coefficients, for a given analytic
representation of $\alpha(t)$.
\end{abstract}
\maketitle
It is often useful to represent the bath response function $\alpha(t)$
in the form:

\begin{equation}
\alpha(t)=\sum_{\bar{K}}^{K}p_{\bar{K}}e^{\Omega_{\bar{K}}t}.\label{eq:bathResponseFunctionIsSumOfExponentials}
\end{equation}

\section{Setting\label{sec:Setting}}

The most popular open quantum system (\textbf{OQS}) model is currently
the Feynman-Vernon model. In the Feynman-Vernon model, the OQS (denoted
by the operator $s$) is coupled linearly to a set of quantum harmonic
oscillators $Q_{k}$:
\begin{alignat}{1}
H & =H_{{\rm OQS}}+H_{{\rm OQS-bath}}+H_{bath}\\
 & =H_{{\rm OQS}}+\sum_{\kappa}c_{\kappa}sQ+\sum_{\kappa}\big(\textrm{\textonehalf}m_{\kappa}\dot{Q_{\kappa}}^{2}+\text{\textonehalf}m_{\kappa}\omega_{\kappa}^{2}Q_{\kappa}^{2}\big)\,.\label{eq:hamiltonianFV}
\end{alignat}
In most models the quantum harmonic oscillators (\textbf{QHO}s) span
a continuous spectrum of frequencies $\omega_{\kappa}$ and the strength
of the coupling between the QHO of frequency $\omega$ and the OQS
is given by the spectral distribution function $J(\omega)$:

\begin{equation}
J(\omega)=\frac{\pi}{2}\sum_{\kappa}\frac{c_{\kappa}^{2}}{m_{\kappa}\omega_{\kappa}}\delta(\omega-\omega_{\kappa})\,.\label{eq:spectralDistribution}
\end{equation}
For the hamiltonian of the Feynman-Vernon model, the bath response
function $\alpha(t)$ is the following integral transform of $J(\omega)$:

\begin{align}
\alpha(t) & =\frac{1}{\pi}\int_{0}^{\infty}J(\omega)\Big(\coth\Big(\frac{\beta\omega\hbar}{2}\Big)\cos(\omega t)-{\rm i}\sin(\omega t)\Big){\rm d}\omega\label{eq:bathResponseFunction}\\
 & =\frac{1}{\pi}\int_{-\infty}^{\infty}\frac{J(\omega)\exp\Big(\frac{\beta\omega\hbar}{2}\Big)}{2\sinh\Big(\frac{\beta\omega\hbar}{2}\Big)}e^{-{\rm i}\omega t}{\rm {\rm d}}\omega,\, J(-\omega)\equiv J(\omega)\label{eq:bathResponseFunctionAsMakri'sFourierTransform}\\
 & =\frac{1}{\pi}\int_{-\infty}^{\infty}\frac{J(\omega)}{1-\exp(-\beta\omega\hbar)}e^{-{\rm i}\omega t}{\rm d}\omega\,,\, J(-\omega)\equiv J(\omega),\label{eq:bathResponseFunctionAsFourierTransformOfBoseFunction}
\end{align}
where, equation \ref{eq:bathResponseFunctionAsFourierTransformOfBoseFunction}
can be written in terms of the Bose-Einstein distribution function
with $x=\beta\omega\hbar$:

\begin{equation}
f^{{\rm Bose-Einstein}}(x)=\frac{1}{1-\exp(-x)}.\label{eq:boseEinsteinFunction}
\end{equation}

\section{Contents\label{sec:contents}}

\subsection{Non-linear least-squares fitting of the bath response function to
the form $\sum_{\bar{K}}^{K}p_{\bar{K}}e^{\Omega_{\bar{K}}t}$}

The goal is to represent $\alpha(t)$ by the series $\sum_{\bar{K}}^{K}p_{\bar{K}}e^{\Omega_{\bar{K}}t}$
as accurately as necessary, with as low as possible a value of $K$.
Given a value of $K$, a non-linear least-squares fitting algorithm
can be used to represent $\alpha(t)$ by that series very accurately.
MATLAB has two such algorithms implemented for easy use: trust region
reflective, and Levenberg-Marquardt. Our MATLAB code allows the user
to choose either method. This least-squares fitting for a chosen value
of $K$ can then be repeated for larger values of $K$ until the resulting
R$\rho$D has converged to satisfaction. 

Since the least-squares fitting algorithm attempts to minimize the
function $|\alpha(t)-\sum_{\bar{K}}^{K}p_{\bar{K}}e^{\Omega_{\bar{K}}t}|$
with respect to the parameters $\{p_{\bar{K}},\Omega_{\bar{K}}\}$,
we call $|\alpha(t)-\sum_{\bar{K}}^{K}p_{\bar{K}}e^{\Omega_{\bar{K}}t}|$
the \textit{objective function}.

\subsubsection{Objective functions and starting values.}

The objective function can be obtained for any spectral distribution
function from equation \ref{eq:bathResponseFunction} by numerical
integration, but for many classes of spectral distribution functions,
it can be obtained much faster by evaluating analytic formulas. For
the first three classes of spectral distribution functions below,
the Padé decomposition scheme first described in \cite{Hu2010,Hu2011}
can already represent $\alpha(t)$ by the series $\sum_{\bar{K}}^{K}p_{\bar{K}}e^{\Omega_{\bar{K}}t}$
very accurately with a small number of terms in the series, so one
can evaluate $\alpha(t)=\sum_{\bar{K}}^{K}p_{\bar{K}}e^{\Omega_{\bar{K}}t}$
with larger and larger values of $K$ until convergence is reached,
using the expressions for the parameters $\{p_{\bar{K}},\Omega_{\bar{K}}\}$
given below. For the fourth class of spectral distribution functions
below, the objective function $\alpha(t)$ can be calculated very
quickly with the closed form analytic expression presented in that
section. There are obviously other spectral distribution functions
for which there are closed form expressions for $\alpha(t)$, or analytic
series representations, but the forms covered in the four subsections
below are enough to represent a wide range of physically relevant
spectral distributions. When no closed form expression or analytic
series is obtainable for the $\alpha(t)$ for a particular form of
$J(\omega)$, one can obtain $\alpha(t)$ from equation \ref{eq:bathResponseFunction}
by numerical integration.

Non-linear least squares fitting algorithms also typically require
\textit{starting values. }For the present case, these are values for
the parameters $\{p_{\bar{K}},\Omega_{\bar{K}}\}$, such that $|\alpha(t)-\sum_{\bar{K}}^{K}p_{\bar{K}}e^{\Omega_{\bar{K}}t}|$
is close to its \textit{global minimum} with respect to $\{p_{\bar{K}},\Omega_{\bar{K}}\}$.
For the first three spectral distribution function forms below, since
the Padé decomposition scheme mentioned above already represents $\alpha(t)$
by the desired series very accurately with a small number of terms,
we can use the analytic expressions for $\{p_{\bar{K}},\Omega_{\bar{K}}\}$
from this scheme (given below) as starting values for the non-linear
least-squares fit. We are not aware of schemes which represent $\alpha(t)$
by the desired series for the final two forms of $J(\omega)$ listed
below, so choosing good starting values for these cases will not be
as simple. The final two subsections below discuss potentially useful
strategies for guessing good starting values for these cases respectively,
although these are not expected to bring $|\alpha(t)-\sum_{\bar{K}}^{K}p_{\bar{K}}e^{\Omega_{\bar{K}}t}|$
as close to its global minimum with respect to $\{p_{\bar{K}},\Omega_{\bar{K}}\}$
as are the starting values derived from the Padé decomposition for
the first three cases.

The expressions given below were first presented in the table 1 in
the apprendix of \cite{Dattani2012b}. 

\medskip{}

\paragraph{\uline{Spectral distribution functions of the generalized Lorentz-Drude/Debye
(gLDD) form}}

\medskip{}

\begin{equation}
J(\omega)=\frac{\omega}{\pi}\sum_{h}^{\invbreve{h}}\bigg(\frac{\lambda{}_{h}\gamma_{h}}{\gamma_{h}^{2}+(\omega-\tilde{\omega}_{h})^{2}}+\frac{\lambda{}_{h}\gamma_{h}}{\gamma_{h}^{2}+(\omega+\tilde{\omega}_{h})^{2}}\bigg)
\end{equation}

\begin{equation}
\alpha(t)=\sum_{\bar{K}}^{K}p_{\bar{K}}e^{\Omega_{\bar{K}}t}
\end{equation}

where:

\medskip{}

\begin{tabular*}{0.95\textwidth}{@{\extracolsep{\fill}}lllc||c}
$p_{\bar{K}}$ & $\Omega_{\bar{K}}$ & Range & \multicolumn{2}{c}{}\tabularnewline
\hline 
\noalign{\vskip\doublerulesep}
\hline 
$\frac{\lambda{}_{\bar{K}}}{\beta\hbar}\left(1-\sum_{\bar{\mathcal{N}}}^{\mathcal{N}}\frac{2\Xi_{\bar{\mathcal{N}}}\Omega_{\bar{K}}^{2}}{\xi_{\bar{\mathcal{N}}}^{2}-\Omega_{\bar{K}}^{2}}\right)+{\rm i}\frac{\lambda{}_{\bar{K}}\Omega_{\bar{K}}}{2}$ & $-\gamma{}_{\bar{K}}+{\rm i}\tilde{\omega}_{\bar{K}}$ & $\bar{K}\in[0,\invbreve{h}]$ & \multicolumn{2}{c}{}\tabularnewline
$\frac{\lambda{}_{\bar{K}}}{\beta\hbar}\left(1-\sum_{\bar{\mathcal{N}}}^{\mathcal{N}}\frac{2\Xi_{\bar{\mathcal{N}}}\Omega_{\bar{K}}^{2}}{\xi_{\bar{\mathcal{N}}}^{2}-\Omega_{\bar{K}}^{2}}\right)+{\rm i}\frac{\lambda{}_{\bar{K}}\Omega_{\bar{K}}}{2}$ & $-\gamma{}_{\bar{K}}-{\rm i}\tilde{\omega}_{\bar{K}}$ & $\bar{K}\in[\invbreve{h}+1,2\invbreve{h}]$ & \multicolumn{2}{l}{$\{\lambda_{\bar{K}},\gamma_{\bar{K}},\tilde{\omega_{\bar{K}}}\}_{\bar{K}=\invbreve{h}+j}^{2\invbreve{h}}\equiv\{\lambda_{\bar{K}},\gamma_{\bar{K}},\tilde{\omega_{\bar{K}}}\}_{\bar{K}=j}$}\tabularnewline
$\frac{4\Xi_{\bar{K}}\xi_{\bar{K}}}{\beta\hbar}\sum_{h}^{\invbreve{h}}\lambda_{h}\gamma_{h}\frac{\big(\xi_{\bar{K}}^{2}-|\Omega_{h}|^{2}\big)}{\lvert\xi_{\bar{K}}^{2}-\Omega_{h}^{2}\rvert^{2}}$ & $-\xi_{\bar{K}}$ & $\bar{K}\in[2\invbreve{h}+1,K]$ & \multicolumn{2}{l}{$\{\lambda_{j},\gamma_{j},\tilde{\omega_{j}}\}_{j=2\invbreve{h}+\bar{K}}^{K}\equiv\{\lambda_{j},\gamma_{j},\tilde{\omega_{j}}\}_{j=\bar{K}}$~~.}\tabularnewline
\end{tabular*}

\smallskip{}

Here $\{\xi_{\bar{\mathcal{N}}},\Xi_{\bar{\mathcal{N}}}\}$ are the
$[\mathcal{N}-1/\mathcal{N}]$ Padé parameters for the Bose-Einstein
distribution function. Expressions for them are given in table \ref{tab:[N-1/N]-Pad=0000E9-parameters}.

\medskip{}

\paragraph{\uline{Spectral distribution functions of the thermally scaled
generalized Lorentz-Drude/Debye (tgLDD) form}}

\medskip{}

\begin{equation}
J(\omega)=\frac{1}{\pi}\tanh\big(\frac{\beta\omega\hbar}{2}\big)\sum_{h}^{\invbreve{h}}\bigg(\frac{\lambda_{h}\gamma_{h}}{\gamma_{h}^{2}+(\omega-\tilde{\omega}_{h})^{2}}+\frac{\lambda_{h}\gamma_{h}}{\gamma_{h}^{2}+(\omega+\tilde{\omega}_{h})^{2}}\bigg)
\end{equation}

\begin{equation}
\alpha(t)=\sum_{\bar{K}}^{K}p_{\bar{K}}e^{\Omega_{\bar{K}}t}
\end{equation}

where:\medskip{}

\begin{tabular*}{0.95\textwidth}{@{\extracolsep{\fill}}lllc||c}
$p_{\bar{K}}$ & $\Omega_{\bar{K}}$ & Range & \multicolumn{2}{l}{}\tabularnewline
\hline 
\noalign{\vskip\doublerulesep}
\hline 
$\frac{\lambda_{h}}{2}+{\rm i}\frac{2\lambda_{h}}{\beta\hbar}\sum_{\bar{\mathcal{N}}}^{\mathcal{N}}\frac{\Xi_{\bar{\mathcal{N}}}\Omega_{h}}{\xi_{\bar{\mathcal{N}}}^{2}-\Omega_{h}^{2}}\,,$ & $-\gamma{}_{\bar{K}}+{\rm i}\tilde{\omega}_{\bar{K}}$ & $\bar{K}\in[0,\invbreve{h}]$ & \multicolumn{2}{c}{}\tabularnewline
$\frac{\lambda_{h}}{2}+{\rm i}\frac{2\lambda_{h}}{\beta\hbar}\sum_{\bar{\mathcal{N}}}^{\mathcal{N}}\frac{\Xi_{\bar{\mathcal{N}}}\Omega_{h}}{\xi_{\bar{\mathcal{N}}}^{2}-\Omega_{h}^{2}}\,,$ & $-\gamma{}_{\bar{K}}-{\rm i}\tilde{\omega}_{\bar{K}}$ & $\bar{K}\in[\invbreve{h}+1,2\invbreve{h}]$ & \multicolumn{2}{l}{$\{\lambda_{\bar{K}},\gamma_{\bar{K}},\tilde{\omega_{\bar{K}}}\}_{\bar{K}=\invbreve{h}+j}^{2\invbreve{h}}\equiv\{\lambda_{\bar{K}},\gamma_{\bar{K}},\tilde{\omega_{\bar{K}}}\}_{\bar{K}=j}$}\tabularnewline
${\rm i}\frac{4\Xi_{\bar{K}}}{\beta\hbar}\sum_{h}^{\invbreve{h}}\tilde{\lambda}_{h}\tilde{\gamma}_{h}\frac{\big(\xi_{\bar{K}}^{2}-|\Omega_{h}|^{2}\big)}{\lvert\xi_{\bar{K}}^{2}-\Omega_{h}^{2}\rvert^{2}}\,,$ & $-\xi_{\bar{K}}$ & $\bar{K}\in[2\invbreve{h}+1,K]$ & \multicolumn{2}{l}{$\{\lambda_{j},\gamma_{j},\tilde{\omega_{j}}\}_{j=2\invbreve{h}+\bar{K}}^{K}\equiv\{\lambda_{j},\gamma_{j},\tilde{\omega_{j}}\}_{j=\bar{K}}$~~.}\tabularnewline
\end{tabular*}

\medskip{}

Here $\{\xi_{\bar{\mathcal{N}}},\Xi_{\bar{\mathcal{N}}}\}$ are the
$[\mathcal{N}-1/\mathcal{N}]$ Padé parameters for the Fermi-Dirac
distribution function. Expressions for them are given in table \ref{tab:[N-1/N]-Pad=0000E9-parameters}.

\medskip{}

\paragraph{\uline{Spectral distribution functions of the Meier-Tanor (MT)
form}}

\medskip{}

\begin{equation}
J(\omega)=\frac{\pi\omega}{2}\sum_{h}^{\invbreve{h}}\frac{\lambda_{h}}{\big(\gamma_{h}^{2}+(\omega+\tilde{\omega}_{h})^{2}\big)\big(\gamma_{h}^{2}+(\omega-\tilde{\omega}_{h})^{2}\big)}
\end{equation}

\begin{equation}
\alpha(t)=\sum_{\bar{K}}^{K}p_{\bar{K}}e^{\Omega_{\bar{K}}t}
\end{equation}

where:

\medskip{}

\begin{tabular*}{0.95\textwidth}{@{\extracolsep{\fill}}lllc||c}
$p_{\bar{K}}$ & $\Omega_{\bar{K}}$ & Range & \multicolumn{2}{l}{}\tabularnewline
\hline 
\noalign{\vskip\doublerulesep}
\hline 
$\frac{\lambda_{h}}{2}+{\rm i}\frac{2\lambda_{h}}{\beta\hbar}\sum_{\bar{\mathcal{N}}}^{\mathcal{N}}\frac{\Xi_{\bar{\mathcal{N}}}\Omega_{h}}{\xi_{\bar{\mathcal{N}}}^{2}-\Omega_{h}^{2}}\,,$ & $-\gamma{}_{\bar{K}}+{\rm i}\tilde{\omega}_{\bar{K}}$ & $\bar{K}\in[0,\invbreve{h}]$ & \multicolumn{2}{c}{}\tabularnewline
$\frac{\lambda_{h}}{2}+{\rm i}\frac{2\lambda_{h}}{\beta\hbar}\sum_{\bar{\mathcal{N}}}^{\mathcal{N}}\frac{\Xi_{\bar{\mathcal{N}}}\Omega_{h}}{\xi_{\bar{\mathcal{N}}}^{2}-\Omega_{h}^{2}}\,,$ & $-\gamma{}_{\bar{K}}-{\rm i}\tilde{\omega}_{\bar{K}}$ & $\bar{K}\in[\invbreve{h}+1,2\invbreve{h}]$ & \multicolumn{2}{l}{$\{\lambda_{\bar{K}},\gamma_{\bar{K}},\tilde{\omega_{\bar{K}}}\}_{\bar{K}=\invbreve{h}+j}^{2\invbreve{h}}\equiv\{\lambda_{\bar{K}},\gamma_{\bar{K}},\tilde{\omega_{\bar{K}}}\}_{\bar{K}=j}$}\tabularnewline
${\rm i}\frac{4\Xi_{\bar{K}}}{\beta\hbar}\sum_{h}^{\invbreve{h}}\tilde{\lambda}_{h}\tilde{\gamma}_{h}\frac{\big(\xi_{\bar{K}}^{2}-|\Omega_{h}|^{2}\big)}{\lvert\xi_{\bar{K}}^{2}-\Omega_{h}^{2}\rvert^{2}}\,,$ & $-\xi_{\bar{K}}$ & $\bar{K}\in[2\invbreve{h}+1,K]$ & \multicolumn{2}{l}{$\{\lambda_{j},\gamma_{j},\tilde{\omega_{j}}\}_{j=2\invbreve{h}+\bar{K}}^{K}\equiv\{\lambda_{j},\gamma_{j},\tilde{\omega_{j}}\}_{j=\bar{K}}$~~.}\tabularnewline
\end{tabular*}

\medskip{}

Here $\{\xi_{\bar{\mathcal{N}}},\Xi_{\bar{\mathcal{N}}}\}$ are the
$[\mathcal{N}-1/\mathcal{N}]$ Padé parameters for the Bose-Einstein
distribution function. Expressions for them are given in table \ref{tab:[N-1/N]-Pad=0000E9-parameters}.

\medskip{}

\paragraph{\uline{\label{par:JwithExponentialCutoff}}}

\medskip{}

\begin{equation}
J(\omega)=A\omega^{s}e^{-\nicefrac{\omega}{\omega_{c}}}
\end{equation}

\begin{equation}
\alpha(t)=A\Re\left(\beta^{-(s+1)}\left(\psi^{(s)}\left(z(t)\right)+\psi^{(s)}\left(z(t)+1\right)\right)\right)+{\rm i}A\Im\left(\frac{\Gamma(s+1)}{(\beta z(t))^{s+1}}\right)\,,\,{\rm where,}
\end{equation}
\begin{equation}
z(t)\equiv\frac{1}{\beta}\big(\frac{1}{\omega_{c}}+{\rm {\rm i}t\big)\,},\,{\rm and,}
\end{equation}

\begin{equation}
\psi^{(s)}(z)\equiv\frac{{\rm d}^{s+1}}{{\rm d}z^{s+1}}\ln\Gamma(z).\label{eq:polyGammaFunction-integralForm}
\end{equation}

For $s\in\mathbb{N}_{0}$ (ie, if $s$ is a non-negative integer),
our definition of $\psi^{(s)}(z)$ in equation \ref{eq:polyGammaFunction-integralForm}
is a well-known representation for the \textit{\textcolor{black}{polygamma
function, }}\textcolor{black}{where }$\Gamma(z)$ is the well-known
gamma function. For these values of $s$, the current versions of
MATLAB and Mathematica have a built in implementation of the polygamma
function that evaluates it in real time. However, there are various
different generalizations of the polygamma function for negative and
non-integer values of $s$ (some popular examples can be found in
\cite{Espinosa2004} and in \cite{Grossman1976}). Mathematica's generalization
of the polygamma function for $s\notin\mathbb{N}_{0}$ \textbf{does
not }in fact give $\psi^{(s)}(z)$ as defined in equation \ref{eq:polyGammaFunction-integralForm}.
Fortunately, Paul Godfrey's implementation of the polygamma function
in his MATLAB function psin(s,z) \textbf{does} give $\psi^{(s)}(z)$
as defined in equation \ref{eq:polyGammaFunction-integralForm} for
all $s\in\mathbb{C}$ with $\Re(s)\ge0$, and the evaluation is computed
in real time. This function can be found on its own \cite{Godfrey2004},
or in Paul Godfrey's bigger special functions package \cite{Godfrey2004a},
both which are available for free at MATLAB Central's File Exchange.

For this spectral distribution function, our recommendation for the
starting values is less straightforward than for the above three cases.
It may be possible to represent $\alpha(t)$ as a sum of complex-weighted
complex exponentials analytically, but unlike the above three cases
where these series were derived from the clever Padé decomposition,
which is mathematically expected to converge with very few terms in
the series, we do not know of any such series that represents $\alpha(t)$
for this $J(\omega)$ such that convergence will be achieved with
very few exponentials. Therefore, we recommend that the objective
function $\alpha(t)$ is calculated using the analytic formula presented
above, and that the starting parameters are chosen based on observing
some of its properties, such as its frequency of oscillation and damping
rate. Guidelines for choosing starting parameters this way are presented
in the subsection below.

\medskip{}

\paragraph{\uline{Other spectral distribution function forms}}

\medskip{}

When $J(\omega)$ is only known in a form such that $\alpha(t)$ is
not easily represented by equation \ref{eq:bathResponseFunctionIsSumOfExponentials}
(as for the case in subsection \ref{par:JwithExponentialCutoff},
or if $J(\omega)$ is only known numerically), it can be much harder
to find good starting values for the non-linear least squares fit.
It might be useful to fit $J(\omega)$ to a form for which good starting
parameters are easy to choose, such as the first three forms presented
above, but if this is not easy, one can calculate the objective function
$\alpha(t)$ by numerical integration of equation \ref{eq:bathResponseFunction},
and as mentioned towards the end of subsection \ref{par:JwithExponentialCutoff},
one can use properties of $\alpha(t)$ as a guide to choose starting
values for $\{p_{\bar{K}},\Omega_{\bar{K}}\}$.

Our recommendation for the starting values for $\{p_{\bar{K}}\}$
is based on the fact that at $t=0$, equation \ref{eq:bathResponseFunctionIsSumOfExponentials}
gives us the relation:

\begin{equation}
\alpha(0)=\sum_{\bar{K}}^{K}p_{\bar{K}}\label{eq:bathReseponseFunctionIsSumOfExponentialsAtTimeZero}
\end{equation}

We recommend to first attempt to fit one term ($p_{1}e^{\Omega_{1}t})$
to the objective function $\alpha(t)$, with $p_{1}=\alpha(0)$ according
to equation \ref{eq:bathReseponseFunctionIsSumOfExponentialsAtTimeZero}
(bear in mind that $\Im(p_{1})$ will then be 0 because $\sin(\omega\cdot0)=0$
will nullify the right side of equation \ref{eq:bathResponseFunction}).
The strating parameter for $\Omega_{1}$ can then be chosen by looking
at a plot of the objective function $\alpha(t)$ and comparing it
to the expressions:

\begin{equation}
\Re(\alpha(t))=p_{1}e^{\Re(\Omega_{1})}\cos(\Im(\Omega_{1})t)\,,
\end{equation}

\begin{equation}
\Im(\alpha(t))=p_{1}e^{\Re(\Omega_{1})}\sin(\Im(\Omega_{1})t)\,\,.
\end{equation}

Based on equation \ref{eq:bathResponseFunction} we see that as the
temperature gets higher, $\Re(\alpha(t))$ becomes more and more different
from $\Im(\alpha(t)$), and therefore more than one damping rate ($\Re(\Omega_{\bar{K}}))$
and more than one angular frequency ($\Im(\Omega_{\bar{K}}))$ will
be needed for a good fit. However, as a crude estimate, we can choose
the starting value of $\Im(\Omega_{1})$ to be an average of the angular
frequencies ($\Omega=\frac{2\pi}{T}$, $T$ $\equiv$period of oscillations)
of the real and imaginary parts of the objective function $\alpha(t)$.
Since the effect of $J(\omega)$ on $\alpha(t)$ will cause the frequency
of oscillations of $\alpha(t)$ to be less and less sinusoidal over
time, we recommend to estimate $T$ of each complex component of $\alpha(t)$
based on the time it takes that component to get to the first quarter
(or half) of its first oscillation.

With this starting value for $\Im(\Omega_{1})$, the starting value
of $\Re(\Omega_{1})$ can then be chosen to be an average of the damping
rates of the real and imaginary parts of the objective function $\alpha(t)$.
For the real part of $\alpha(t)$, the damping rate $\Re(\Omega_{1,\Re})$
can be estimated by observing the time $t_{\alphaRealFirstMin}$ at
which $\Re(\alpha(t))$ attains its first minimum $\alphaRealFirstMin$,
and then solving the equation:

\begin{eqnarray}
\alphaRealFirstMin & = & \exp\left(\Re(\Omega_{1,\Re})t_{\alphaRealFirstMin}\right)\cos(\Im(\Omega_{1})t_{\alphaRealFirstMin})\,,\,\textrm{whose solution is}\\
\Re(\Omega_{1,\Re}) & = & \frac{\ln\left(\frac{\alphaRealFirstMin}{p_{1}\cos(\Im(\Omega_{1})t_{\alphaRealFirstMin})}\right)}{t_{\alphaRealFirstMin}}
\end{eqnarray}
which is an estimate of $\Re(\Omega_{1,\Re})$. 

Similarly, the damping rate of the imaginary part of $\alpha(t)$,
which we denote by $\Re(\Omega_{1,\Im})$, can be estimated by%
\footnote{Since this expression uses the amplitude $p_{1}$ which is derived
from $\Re(\alpha(t)),$ and $\Re(\alpha(t))$ is expected to deviate
more and more from $\Im(\alpha(t))$ as the temperature is increased
(based on equation \ref{eq:bathResponseFunction}), this estimate
is expected to be best at low temperatures.%
}:

\begin{equation}
\Re(\Omega_{1,\Im})=\frac{\ln\left(\frac{\alphaRealFirstMin}{p_{1}\sin(\Im(\Omega_{1})t_{\alphaRealFirstMin})}\right)}{t_{\alphaRealFirstMin}}\,.
\end{equation}

$\Re(\Omega_{1})$ is then estimated as an average of $\Re(\Omega_{1,\Re})$
and $\Im(\Omega_{1,\Im})$.

We can then fit the one term function $p_{1}e^{\Omega_{1}t}$ to the
objective function, and then use the resulting fitted values of $p_{1}$
and $\Omega_{1}$ as starting values for a fit of the two term series
$p_{1}e^{\Omega_{1}t}+p_{2}e^{\Omega_{2}t}$ to the objective function.
Since $p_{2}$ and $\Omega_{2}$ are expected to be of the same order
of magnitude as $p_{1}$and $\Omega_{1}$ respectively, we can multiply
$p_{1}$and $\Omega_{1}$ by random numbers near 1 in order to get
crude estimates of suitable starting values for $p_{2}$and $\Omega_{2}$
respectively. In our MATLAB program, these random numbers are obtained
using (1+RANDN) in MATLAB, rather than RAND or RANDN, so that these
random numbers are more likely to be close to 1 rather than to 0.
The resulting values of $p_{1},\Omega_{1},p_{2}$ and $\Omega_{2}$
from this fit can then be used as starting values for a three term
fit, with starting values for $p_{3}$ and $\Omega_{3}$ chosen as
$p_{2}$ and $\Omega_{2}$ were for the two term fit. 

Fits to series with larger numbers of terms can be done in a similar
way, although when there are many terms in the series, it may be more
appropriate (based on equation \ref{eq:bathReseponseFunctionIsSumOfExponentialsAtTimeZero})
for the starting value of $p_{1}$ to be $\nicefrac{\alpha(0)}{K}$
instead of just $\alpha(0)$, though the other starting values would
likely also have to be adjusted, which would not be straightforward.

\subsubsection{Note about implementation: Constraints, scaling \& step sizes, and
weights}

Putting constraints on the fitting parameters can help speed up the
fits, and can prevent the fitting program from getting lost in a far
from optimal local minimum, or the fitted values form becoming grossly
unphysical. For example, it helps to implement the constraint $\Re(\Omega_{\bar{K}})<0$,
so that $\alpha(t)$ is not likely to diverge. Adding this as a constraint
to the fit will prevent the fitting program from bothering to try
values that we know are expected to give bad results, and will therefore
speed up the fitting caluclation, and could potentially prevent the
fitting program from getting `lost' in a region far from the desired
global minimum. We have implemented these constraints in BATHFIT's
fitting routine. When $K$ is small, it may also help to implement
the constraint $p_{1}>0$, since according to equation \ref{eq:bathReseponseFunctionIsSumOfExponentialsAtTimeZero}
and \ref{eq:bathResponseFunction}, if $K=1$, $p_{1}>0$. 

Since MATLAB's fitting routines are most easily implemented when the
step sizes are the same size, we scale the objective function's range
to be between 0 and 1, and its domain is mapped to $t\in[0,1]$, which
significantly helps preventing the fitting program from getting lost
in regions far from the desired global minimum. After the fit is complete,
we then scale $\{p_{\bar{K}},\Omega_{\bar{K}}\}$ back to SI units.

We also provide the user with the option of assigning weights to each
datapoint of the objective function. For example, if the user requires
the first 100fs of $\alpha(t)$ to be represented very accurately,
and does not require $\alpha(t)$ to be represented very accurately
after 500fs, the user can assign weights accordingly. Often this can
also help to get a graphically better fit, since the local minimum
found by BATHFIT won't always be the closest to the global minimum,
or even if the global minimum is attained, it may have been influenced
too strongly by certain parts of $\alpha(t)$ which are not so important.

\subsection{Analytic coefficients for the discretized influence functional}

Once $\alpha(t)$ is represented in the form of equation \ref{eq:bathResponseFunctionIsSumOfExponentials},
the \textbf{DIF}s (discretized influence functional coefficients,
which are required if using a Feynman integral to calculate the $R\rho D$
of open quantum systems modeled by the Feynman-Vernon model), can
be calculated very quickly by the analytic formulas which were presented
in \cite{Dattani2012b}. These formulas are listed again below, along
with analogous formulas for spectral distribution functions of the
form $J(\omega)=A\omega^{s}e^{-(\nicefrac{\omega}{\omega_{c}})}$.
For this latter form for the spectral distribution function, $\alpha(t)$
can once again be fitted to the form of equation \ref{eq:bathResponseFunctionIsSumOfExponentials},
and the DIFs can therefore be calculated with the first set of formulas
below, but since we have an exact analytic form for $\alpha(t)$ that
can easily be integrated with respect to $t$, we have also presented
analytic forms for the DIFs for this form of $J(\omega)$, which do
not require $\alpha(t)$ to first be fitted to the form of equation
\ref{eq:bathResponseFunctionIsSumOfExponentials}.

\subsubsection{Spectral distribution functions with bath response functions of the
form $\alpha(t)=\sum_{\bar{K}}^{K}p_{\bar{K}}e^{\Omega_{\bar{K}}t}$}

\paragraph{\uline{Trotter splitting}}

\begin{alignat}{1}
\eta_{kk^{\prime}} & =4\sum_{\bar{K}=1}^{K}\frac{p_{\bar{K}}}{\Omega_{\bar{K}}^{2}}\sinh^{2}(\nicefrac{\Omega_{\bar{K}}\Delta t}{2})e^{\Omega_{\bar{K}}(k-k^{\prime})\Delta t}\,\,,\,0\le k^{\prime}<k\le N\,\,,\,\,{\rm and}\label{eq:etaKK'whenAlphaIsSumOfExponentials}\\
\eta_{kk} & =2\sum_{\bar{K}=1}^{K}\frac{p_{\bar{K}}}{\Omega_{\bar{K}}^{2}}\bigg(\sinh(\nicefrac{\Omega_{\bar{K}}\Delta t}{2})e^{\nicefrac{\Omega_{\bar{K}}\Delta t}{2}}-\frac{1}{2}\Omega_{\bar{K}}\Delta t\bigg)\,\,,\,0\le k\le N\,.\label{eq:etaKKwhenAlphaIsSumOfExponentials}
\end{alignat}

\paragraph{\uline{Strang splitting}}

\begin{alignat}{1}
\eta_{N0} & =4\sum_{j=1}^{K}\frac{p_{\bar{K}}}{\Omega_{\bar{K}}^{2}}e^{\Omega_{\bar{K}}(t-\nicefrac{\Delta t}{2})}\left(\sinh^{2}(\nicefrac{\Omega_{\bar{K}}\Delta t}{4})\right)\,\,,\label{eq:etaN0whenAlphaIsSumOfExponentials}\\
\eta_{00} & =\eta_{NN}=2\sum_{j=1}^{K}\frac{p_{\bar{K}}}{\Omega_{\bar{K}}^{2}}\left(e^{\nicefrac{\Omega_{\bar{K}}\Delta t}{4}}\sinh(\nicefrac{\Omega_{\bar{K}}\Delta t}{4})-\nicefrac{\Delta t\Omega_{j}}{4}\right)\,\,,\label{eq:eta00whenAlphaIsSumOfExponentials}\\
\eta_{k0} & =4\sum_{j=1}^{K}\frac{p_{\bar{K}}}{\Omega_{\bar{K}}^{2}}\sinh(\nicefrac{\Omega_{\bar{K}}\Delta t}{2})\sinh\left(\nicefrac{\Omega_{\bar{K}}\Delta t}{4}\right)e^{\Omega_{\bar{K}}(k\Delta t-\nicefrac{\Delta t}{4})}\,\,,\label{eq:etaK0whenAlphaIsSumOfExponentials}\\
\eta_{Nk} & =4\sum_{j=1}^{K}\frac{p_{\bar{K}}}{\Omega_{\bar{K}}^{2}}\sinh(\nicefrac{\Omega_{\bar{K}}\Delta t}{2})\sinh\left(\nicefrac{\Omega_{\bar{K}}\Delta t}{4}\right)e^{\Omega_{\bar{K}}(t-k\Delta t-\nicefrac{\Delta t}{4})}\,\,.\label{eq:etaNKwhenAlphaIsSumOfExponentials}
\end{alignat}

\subsubsection{Spectral distribution functions with exponential cut-offs\label{sub:DIFsForJwithExponentialCutoff}}

\subsubsection{QUAPI}

When the bath is nearly adiabatic, it is helpful to rewrite equation
\ref{eq:hamiltonianFV} as \cite{Leggett1987,1992Makri}:

\begin{alignat}{1}
H & =H_{{\rm OQS}}-H_{{\rm displacement}}+H_{{\rm OQS-bath}}+H_{{\rm bath}}+H_{{\rm displacement}}\label{eq:counterTermDisplacementGeneralBath}\\
 & =H_{{\rm OQS}}-\sum_{\kappa}\frac{c_{\kappa}^{2}s^{2}}{2m_{\kappa}\omega_{\kappa}^{2}}+\sum_{\kappa}c_{\kappa}sQ+\sum_{\kappa}\big(\textrm{\textonehalf}m_{\kappa}\dot{Q_{\kappa}}^{2}+\text{\textonehalf}m_{\kappa}\omega_{\kappa}^{2}Q_{\kappa}^{2}\big)+\sum_{\kappa}\frac{c_{\kappa}^{2}s^{2}}{2m_{\kappa}\omega_{\kappa}^{2}}\label{eq:counterTermDisplacementForFeynmanVernonBath}\\
 & \equiv H_{{\rm OQS,displaced}}+\sum_{\kappa}c_{\kappa}sQ+H_{{\rm bath,displaced}}.\label{eq:counterTermDisplacementForFeynmanVernonBathInCompactForm}
\end{alignat}
 $H_{{\rm displacement}}$ is called the {}``counter term'', and
can also be represented in terms of the spectral distribution function
by recognizing that when the QHOs span a continuous spectrum of frequencies
$\omega_{\kappa}$ , we have the relation (remembering equation \ref{eq:spectralDistribution}):

\begin{equation}
\sum_{\kappa}\frac{c_{\kappa}^{2}}{2m_{\kappa}\omega_{\kappa}^{2}}=\frac{1}{\pi}\int_{0}^{\infty}\frac{J(\omega)}{\omega}{\rm d}\omega\,.
\end{equation}

A Feynman integral used to calcualte the $R\rho D$ for an OQS denoted
by the hamiltonian $H_{{\rm OQS,displaced}}$, is called a QUAPI,
which stands for \textbf{Qu}asi\textbf{-A}diabatic\textbf{ P}ropagator
Feynman%
\footnote{The term `path integral' is used more commonly than `Feynamn integral'
here, but this term is ambiguous. Currently, the first result on the
search engine at www.google.com, when the search query `path integral'
is entered, is a Wikipedia page that currently links to three different
meanings of the word `path integral': (1) line integral, (2) functional
integration, and (3) path integral formulation. Only the third of
these is unambiguously the Feynman integral discussed in this paper.
The `line integral' is an integral over a path, rather than over a
set of paths; and the term `functional integral' can refer to at least
three types of functional integrals: (1) the Wiener integral, (2)
the Lévy integral, and (3) the Feynman integral.%
} \textbf{I}ntegral. For QUAPI calculations, all $\eta$ coefficients
remain the same as for a feynman integral used to calculate the $R\rho D$
for an OQS whose hamiltonian has not been displaced according to equations
\ref{eq:counterTermDisplacementGeneralBath}\ref{eq:counterTermDisplacementForFeynmanVernonBath}\ref{eq:counterTermDisplacementForFeynmanVernonBathInCompactForm};
except for the $\eta_{kk^{\prime}}$ coefficients when $k=k^{\prime}$
($\forall k\in[0,N]$), which change to (see supplementary material
of \cite{Dattani2012b}): 

\begin{flushleft}
\begin{alignat}{1}
\eta_{kk}^{{\rm QUAPI}} & =\eta_{kk}+\frac{{\rm i}\Delta t}{\hbar\pi}\int_{0}^{\infty}\frac{J(\omega)}{\omega}{\rm d}\omega\,,\,{\rm for}\, k\in[0,N]\label{eq:QUAPItermInInfluenceFunctionalWRTspectralDistribution}\\
 & \equiv\eta_{kk}+\frac{{\rm i}\Delta t}{\hbar\pi}\lambda\,,
\end{alignat}
 where in the last line we have defined the {}``bath reorganization
energy'' by $\lambda$. An analytic expression exists for $\lambda$
for most forms of $J(\omega)$ presented in this paper:
\par\end{flushleft}

\begin{center}
\begin{tabular*}{0.95\textwidth}{@{\extracolsep{\fill}}ccc}
 & $J(\omega)$ & $\lambda$\tabularnewline
\hline 
\noalign{\vskip\doublerulesep}
\hline 
\textbf{gLDD } & $J(\omega)=\frac{\omega}{\pi}\sum_{h}^{\invbreve{h}}\bigg(\frac{\lambda{}_{h}\gamma_{h}}{\gamma_{h}^{2}+(\omega-\tilde{\omega}_{h})^{2}}+\frac{\lambda{}_{h}\gamma_{h}}{\gamma_{h}^{2}+(\omega+\tilde{\omega}_{h})^{2}}\bigg)$ & $\sum_{h}^{\invbreve{h}}\lambda_{h}$\tabularnewline
\textbf{MT} & $\frac{\pi\omega}{2}\sum_{h}^{\invbreve{h}}\frac{\lambda_{h}}{\big(\gamma_{h}^{2}+(\omega+\tilde{\omega}_{h})^{2}\big)\big(\gamma_{h}^{2}+(\omega-\tilde{\omega}_{h})^{2}\big)}$ & $\frac{\pi^{2}}{8\gamma(\gamma^{2}+\omega_{0}^{2})}\sum_{h}^{\invbreve{h}}\lambda_{h}$\tabularnewline
 & {\footnotesize $A\omega^{s}e^{-(\nicefrac{\omega}{\omega_{c}})^{q}}\, s>0\,,\,\omega_{c}>0\,,\, q>0$} & $\frac{A}{q}\omega_{c}^{s}\Gamma(\frac{s}{q})$\tabularnewline
\end{tabular*}
\par\end{center}

\section{Appendix}

\begin{table}[H]
\caption{$[\mathcal{N}-1,\mathcal{N}]$ Padé parameters for the Bose-Einstein
and Fermi-Dirac distribution functions (first presented in \cite{Hu2010}
and in more detail in \cite{Hu2011}). $\{\Xi_{\bar{\mathcal{N}}},\xi_{\bar{\mathcal{N}}}\}$
can be calculated easily for all $\bar{\mathcal{N}}$, for arbitrary
values of $\bar{\mathcal{N}}$ in our open source MATLAB program that
supplements this paper. The matrix $\Lambda$ is a 2$\mathcal{N}$$\times$2$\mathcal{N}$,
and $\tilde{\Lambda}$ is a 2$\mathcal{N}$-1$\times$2$\mathcal{N}$-1.
The indices $\bar{\mathcal{N}}$ run from 1 to half the number of
non-zero eigenvalues of the corresponding matrix. %
\begin{lyxgreyedout}
The number of non-zero eigenvalues will always be even, because the
eigenvalues of these particular matrices come in pairs, for example
($\xi_{\bar{\mathcal{N}}},-\xi_{\bar{\mathcal{N}}}$); and for matrices
with an odd number of eigenvalues, the unpaired eigenvalue will always
be 0.%
\end{lyxgreyedout}
\label{tab:[N-1/N]-Pad=0000E9-parameters}}

\centering{}%
\begin{tabular*}{0.6\textwidth}{@{\extracolsep{\fill}}cc}
\multicolumn{1}{c}{$\{\pm\xi_{\bar{\mathcal{N}}}\}=\frac{2}{\text{\ensuremath{\beta\hbar\cdot}eigenvalues}(\Lambda)}$} & $\{\pm\zeta_{\bar{\mathcal{N}}}\}=\frac{2}{\beta\hbar\cdot{\rm eigenvalues}(\tilde{\Lambda})}$\tabularnewline
\hline 
\hline 
Bose-Einstein  & Fermi-Dirac\tabularnewline
$\Lambda_{mn}=\frac{\delta_{m,n\pm1}}{\sqrt{(2m+1)(2n+1)}}$  & $\Lambda_{mn}=\frac{\delta_{m,n\pm1}}{\sqrt{(2m-1)(2n-1)}}$\tabularnewline
$\tilde{\Lambda}_{mn}=\frac{\delta_{m,n\pm1}}{\sqrt{(2m+3)(2n+3)}}$  & $\tilde{\Lambda}_{mn}=\frac{\delta_{m,n\pm1}}{\sqrt{(2m+1)(2n+1)}}$\tabularnewline
$\Xi_{\bar{\mathcal{N}}}=(\mathcal{N}^{2}+\frac{3}{2}\mathcal{N})\frac{\prod_{\bar{\bar{\mathcal{N}}}=1}^{\mathcal{N}-1}(\zeta_{\bar{\bar{\mathcal{N}}}}^{2}-\xi_{\bar{\mathcal{N}}}^{2})}{\prod_{\bar{\bar{\mathcal{N}}}\neq\bar{\mathcal{N}}}^{\mathcal{N}}(\xi_{\bar{\bar{\mathcal{N}}}}^{2}-\xi_{\bar{\mathcal{N}}}^{2})}$  & $\Xi_{\bar{\mathcal{N}}}=(\mathcal{N}^{2}+\frac{1}{2}\mathcal{N})\frac{\prod_{\bar{\bar{\mathcal{N}}}=1}^{\mathcal{N}-1}(\zeta_{\bar{\bar{\mathcal{N}}}}^{2}-\xi_{\mathcal{\bar{N}}}^{2})}{\prod_{\bar{\bar{\mathcal{N}}}\neq\bar{\mathcal{N}}}^{\mathcal{N}}(\xi_{\bar{\bar{\mathcal{N}}}}^{2}-\xi_{\bar{\mathcal{N}}}^{2})}$\tabularnewline
\end{tabular*}
\end{table}

\section{Acknowledgements}

N.S.D. thanks the Clarendon Fund and the NSERC/CRSNG of/du Canada
for financial support. F.A.P. thanks the Leverhulme Trust for financial
support.

\bibliographystyle{plain}

\end{document}